\input amstex
\documentstyle{amsppt}

\NoRunningHeads
\newcount\qcts\newcount\qcta\qcta=1\newcount\qcteq\newcount\qcthead
\def\sethead#1{\global\qcts=0\global\advance\qcthead by\number\qcta
       \immediate\write3{
    \string\def\string#1\string{\number\qcthead\string}}}
\def\setref#1{\global\qcteq=0\global\advance\qcts by\number\qcta
       \immediate\write3{
    \string\def\string#1\string{\number\qcthead.\number\qcts\string}}}
\def\seteqn#1{\global\advance\qcteq by\number\qcta\immediate\write3{
    \string\def\string#1\string{\number\qcthead.\number\qcts.\eqnpbg\string}}}
\def\eqnpbg{\ifnum\qcteq=1 a\else\ifnum\qcteq=2 b\else
   \ifnum\qcteq=3 c\else\ifnum\qcteq=4 d\else
   \ifnum\qcteq=5 e\else\ifnum\qcteq=6 f\else\ifnum\qcteq=7 g\else
   \ifnum\qcteq=8 h\else\ifnum\qcteq=9 i\else
   \ifnum\qcteq=10 j\else\ifnum\qcteq=11 k\else\ifnum\qcteq=12 l\else
   \ifnum\qcteq=13 m\else\ifnum\qcteq=14 n\else\ifnum\qcteq=15 o\else
   \ifnum\qcteq=16 p\else\ifnum\qcteq=17 q\else
   \ifnum\qcteq=18 r\else\ifnum\qcteq=19 s\else\ifnum\qcteq=20 t\else
   \ifnum\qcteq=21u\else\ifnum\qcteq=22 v\else
   \ifnum\qcteq=23 w\else\ifnum\qcteq=24 x\else
   \ifnum\qcteq=25 y\else\ifnum\qcteq=26 z\else *
    \fi\fi\fi\fi\fi\fi\fi\fi\fi\fi\fi\fi\fi\fi\fi\fi\fi\fi\fi\fi\fi\fi\fi\fi\fi\fi}
\newcount\qct\newcount\qcta\qct=0\qcta=1
\def\trace{\operatorname{Tr}}
\def\cupd{\buildrel\ldotp\over\sqcup}
\def\birdy#1{\textstyle\frac\partial{\partial\varrho}\{#1\}|_{\varrho=0}}
\def\pbgkey#1{\key{#1}\global\advance\qct by\number\qcta
       \immediate\write3{\string\def\string#1\string{\number\qct\string}}}
\def\bork{\smallskip\qquad\qquad}
\def\BORK{\smallbreak\qquad}
\def\nmonth{\ifcase\month\ \or January\or
   February\or March\or April\or May\or June\or July\or August\or
   September\or October\or November\else December\fi}
\def\birdy#1{\textstyle\frac\partial{\partial\varrho}\{#1\}|_{\varrho=0}}
\def\nmonth{\ifcase\month\ \or January\or
   February\or March\or April\or May\or June\or July\or August\or
   September\or October\or November\else December\fi}
\pageno=1
\topmatter
\title Heat Trace Asymptotics of a time dependent process\endtitle
\author Peter Gilkey, Klaus Kirsten, 
and JeongHyeong Park\endauthor
\address Peter Gilkey, 
Mathematics  Department, University of Oregon, Eugene Oregon
97403 USA,
email: gilkey\@darkwing.uoregon.edu\endaddress
\address Klaus Kirsten, Department of Physics and Astronomy, The University of
     Manchester, Oxford Road, Manchester UK M13 9PL UK,
     email: klaus\@a13.ph.man.ac.uk\endaddress
\address JeongHyeong Park, Math Dept, Honam University,
      Seobongdong 59, Kwangsanku, Kwangju, 506-090 South Korea,
      email: jhpark\@honam.honam.ac.kr\endaddress
\rightheadtext{Gilkey-Kirsten-Park}
\leftheadtext{Heat Trace Asymptotics}
\abstract We study the heat trace asymptotics defined by a time dependent family
of operators of Laplace type 
which naturally appears for time dependent metrics. 
\endabstract
\endtopmatter

\parindent0mm
MSC number: 58G25, 35P05.

Key words: Spectral analysis, spectral geometry, Laplacian, heat equation
asymptotics.

\immediate \def\qctSA{1}
 \def\arefa{1.1}
 \def\arefaa{1.1.a}
 \def\arefab{1.1.b}
 \def\arefac{1.1.c}
 \def\arefaC{1.1.d}
 \def\arefad{1.1.e}
 \def\arefae{1.1.f}
 \def\arefaf{1.1.g}
 \def\arefag{1.1.h}
 \def\arefb{1.2}
 \def\arefc{1.3}
 \def\qctSB{2}
 \def\brefa{2.1}
 \def\brefba{2.2}
 \def\brefb{2.3}
 \def\brefc{2.4}
 \def\qctSC{3}
 \def\crefa{3.1}
 \def\crefb{3.2}
 \def\qctSD{4}
 \def\drefa{4.1}
 \def\drefb{4.2}
 \def\qctSE{5}
 \def\erefa{5.1}
 \def\erefb{5.2}
 \def\qctSF{6}
 \def\frefa{6.1}
\def\refBiDa{1}
\def\refBrGi{2}
\def\refBrGiKiVa{3}
\def\refGilkey{4}
\def\refGilkeya{5}
\def\refGrubb{6}
\def\refKCD{7}
\def\refM{8}
\def\refMD{9}
\def\refWeyl{10}
\sethead\qctSA\head\S\qctSA\ Introduction\endhead
\setref\arefa
Let $M$ be an $m$ dimensional compact Riemannian manifold with smooth boundary, let
$V$ be a smooth vector bundle over $M$, and let
$D:C^\infty(V)\rightarrow C^\infty(V)$ be an operator of Laplace type whose coefficients
are independent of the parameter $t$; such an operator is said to be static. There is a
canonical connection $\nabla$ on $V$ and a canonical endomorphism $E$ of $V$ so
\seteqn\arefaa
$$D=-\{\trace(\nabla^2)+E\}.\tag\arefaa$$
Let $x=(x_1,...,x_m)$ be a system of local coordinates on $M$. We adopt
the Einstein convention and sum over repeated indices.
Fix a local frame for $V$ and expand:
$$ds_M^2=g_{\mu\nu}dx^\mu\circ dx^\nu\text{ and }
  D=-(g^{\mu\nu}\partial_\mu\partial_\nu+A^\mu\partial_\mu+B)$$
where $A$ and $B$ are local sections of $TM\otimes\text{End}(V)$ and $\text{End}(V)$.
Let $I_V$ be the identity map on $V$. The connection $1$ form $\omega$ of $\nabla$ and
the endomorphism
$E$ appearing in equation (\arefaa) are given by
\seteqn\arefab
$$\eqalign{&\textstyle\omega_\delta=\frac12g_{\nu\delta}(A^\nu+g^{\mu\sigma} 
    \Gamma_{\mu\sigma}{}^\nu I_V)\text{ and}\cr
   &E=B-g^{\nu\mu}(\partial_\nu\omega_\mu 
    +\omega_\nu\omega_\mu-\omega_\sigma\Gamma_{\nu\mu}{}^\sigma);}\tag\arefab$$
see \cite{\refGilkey} for details. Let `;' denote multiple covariant differentiation; we
use the Levi-Civita connection on $M$ and the connection of equation (\arefab) determined
by $D$ to differentiate tensors of all types. If $\Cal{D}$ is a time dependent family of
operators of Laplace type, then we expand $\Cal{D}$ in a Taylor series expansion in $t$ to
write $\Cal{D}$ invariantly in the form:
\seteqn\arefac
$$\Cal{D}u:=Du+\textstyle\sum_{r>0}t^r\{\Cal{G}_{r,}{}^{ij}u_{;ij}
        +\Cal{F}_{r,}{}^iu_{;i}+\Cal{E}_ru\}.\tag\arefac$$
This setting appears most naturally when defining an adiabatic vacuum 
in quantum field theory in curved spacetime \cite{\refBiDa}. If the spacetime
is slowly varying, then the time dependent metric describing the cosmological 
evolution can be expanded in a Taylor series with respect to $t$. The 
index $r$ in this situation is then related to the adiabatic order.
 
Near the boundary, let indices $a,b,...$ range from $1$ through $m-1$ and index a local
orthonormal frame for the boundary; let $e_m$ denote the inward unit normal. We assume
given a decomposition of the boundary  $\partial M=C_{\Cal{N}}\cupd C_{\Cal{D}}$ as the
disjoint union of closed sets - we permit $C_{\Cal{N}}$ or $C_{\Cal{D}}$ to be empty.  Let
\seteqn\arefaC
$$\Cal{B}u:=u|_{C_{\Cal D}}\oplus(u_{;m}+Su+t(T^au_{;a}+S_1u))|_{C_{\Cal N}}\tag\arefaC$$
define the boundary conditions; 
we can treat both 
Robin and Dirichlet boundary conditions with this formalism. In the following
we shall let $\Cal{B}_0$ be the static (i.e. time
independent) part of the boundary  condition; 
$\Cal{B}_0u:=u|_{C_\Cal D}\oplus(u_{;m}+Su)|_{C_{\Cal N}}$. 
The reason for including a
time-dependence in the boundary  condition comes e.g. from considerations of the dynamical
Casimir effect; it takes the form given in (\arefaC) for slowly moving boundaries.  Here
we included only linear powers of $t$ because higher orders do not enter into the
asymptotic terms we are going to calculate.  Note that by multiplying $\Cal{B}$ by
$(1+T^m)^{-1}$, we can take $T^m=0$.

If
$\phi$ is the initial temperature distribution, the subsequent temperature distribution
$u_\phi(t,x)$ is determined by the equations:
\seteqn\arefad
$$(\partial_t+\Cal{D})u_\phi(t,x)=0,\ \Cal{B}u=0,\text{ and }u_\phi(0,x)=\phi.\tag\arefad$$
Let $\Cal{K}:\phi\rightarrow u_\phi$ be the fundamental solution of the heat equation. If
$\Cal{D}$ and $\Cal{B}$ are static, then $\Cal{K}=e^{-tD_\Cal{B}}$. Let $\nu_M$ be the
Riemannian measure on $M$. There exists a smooth endomorphism valued kernel
$K(t,x,\bar x,\Cal{D},\Cal{B}):V_{\bar x}\rightarrow V_x$  so
$$u_\phi(t,x)=(\Cal{K}\phi)(t,x)
  =\textstyle\int_MK(t,x,\bar x,\Cal{D},\Cal{B})\phi(\bar x)d\bar\nu_M.$$
For fixed $t$, the operator $\Cal{K}(t):\phi\rightarrow\phi(t,\cdot)$ is of trace class.
We let
\seteqn\arefae
$$a(f,\Cal{D},\Cal{B})(t):=
     \trace_{L^2}(f\Cal{K}(t))
   =\textstyle\int_Mf(x)\trace_{V_x}(K(t,x,x,\Cal{D},\Cal{B}))d\nu_M.\tag\arefae$$ 
The function $f\in C^\infty(M)$ is introduced as a localizing or smearing function. As
$t\downarrow0$, one can extend the analysis of \cite{\refGrubb} from the static setting to
show that there is a complete asymptotic expansion of the form
\seteqn\arefaf
$$a(f,\Cal{D},\Cal{B})(t)\sim
\textstyle\sum_{n\ge0}a_n(f,\Cal{D},\Cal{B})t^{(n-m)/2}.\tag\arefaf$$
The asymptotic coefficients $a_n(f,\Cal{D},\Cal{B})$ form the focus of
our study. We may decompose $a_n$ into an interior and a boundary contribution:
$$a_n(f,\Cal{D},\Cal{B})=a_n^M(f,\Cal{D})+a_n^{\partial M}(f,\Cal{D},\Cal{B}).$$
The interior invariants vanish if $n$ is odd and do not depend on the boundary condition;
the boundary invariants are generically non-zero for all
$n$. Let
$N^\mu(f)$ denote the
$\mu^{th}$ covariant derivative of the smearing function $f$ with respect to $e_m$. There
exist locally computable invariants $a_n^M(x,\Cal{D})$ and
$a_{n,\mu}^{\partial M}(y,\Cal{D},\Cal{B})$ defined for interior points $x\in M$ and
boundary points
$y\in\partial M$ so that
\seteqn\arefag
$$\eqalign{
&a_n^M(f,\Cal{D})=\textstyle\int_Mf(x)a_n^M(x,\Cal{D})d\nu_M,\text{ and}\cr
&a_n^{\partial M}(f,\Cal{D},\Cal{B})=
     \textstyle\sum_\mu\int_{\partial M}N^\mu(f)
     a_{n,\mu}^{\partial M}(y,\Cal{D},\Cal{B})d\nu_{\partial M}.}\tag\arefag$$

If $\Cal{D}$ and $\Cal{B}$ are static, then these are the heat trace asymptotics which
have been studied in many contexts previously; $a(1,D,\Cal{B})=\trace_{L^2}e^{-tD_{\Cal
B}}$. Let $R_{ijkl}$ be the components of the curvature tensor defined by the Levi-Civita
connection and let $\Omega_{ij}$ be the components of the curvature endomorphism defined by
the auxiliary connection $\nabla$ on $V$. We do not introduce explicit bundle indices for
$\Omega_{ij}$ and $E$. Let $L_{aa}$ be the second fundamental form. Let `:' denote multiple
covariant differentiation with respect to the Levi-Civita connection of the boundary and
the connection defined by $D$. We refer to \cite{\refBrGi} and \cite{\refGilkey} for the
proof of the following result for static $D$; see also related work
\cite{\refBrGiKiVa, \refKCD, \refM, \refMD}.
\setref\arefb
\proclaim{\arefb\ Theorem}
\roster
\smallskip\item $a_0^M(f,D)=(4\pi)^{-m/2}\int_Mf\trace(I_V)d\nu_M$.
\smallskip\item $a_2^M(f,D)=(4\pi)^{-m/2}\textstyle{1\over6}
     \int_Mf\trace(R_{ijji}I_V+6E)d\nu_M$.
\smallskip\item $a_4^M(f,D)=(4\pi)^{-m/2}\textstyle{1\over360}
    \int_Mf\trace\{60E_{;kk}+60R_{ijji}E
    +180E^2+30\Omega_{ij}\Omega_{ij}$
\par\qquad $+(12R_{ijji;kk}+5R_{ijji}R_{kllk}-2R_{ijki}R_{ljkl}
     +2R_{ijkl}R_{ijkl})I_V\}d\nu_M$.
\smallskip\item $a_0^{\partial M}(f,D,\Cal{B})=0$.
\smallskip\item $a_1^{\partial M}(f,D,\Cal{B})=-(4\pi)^{(1-m)/2}\frac14
     \int_{C_D}f\trace(I_V)d\nu_{\partial M}$
\par$+
   (4\pi)^{(1-m)/2}\frac14\int_{C_N}f\trace(I_V)d\nu_{\partial M}$.
\smallskip\item $a_2^{\partial M}(f,D,\Cal{B})=(4\pi)^{-m/2}\frac16\int_{C_D}
\trace\{2fL_{aa}I_V-3f_{;m}I_V\}d\nu_{\partial M}$
\par
$+(4\pi)^{-m/2}\frac16\int_{C_N}\trace\{f(2L_{aa}I_V+12S)+3f_{;m}I_V\}d\nu_{\partial M}$.
\smallskip\item $a_3^{\partial M}(f,D,\Cal{B})=-(4\pi)^{(1-m)/2}\frac1{384}
\int_{C_D}\trace\{96fE+f(16R_{ijji}$
\par$-8R_{amma}+7L_{aa}L_{bb}-10L_{ab}L_{ab})I_V-30f_{;m}L_{aa}I_V+24f_{;mm}I_V\}
    d\nu_{\partial M}$
\par
$+(4\pi)^{(1-m)/2}\frac1{384}\int_{C_N}\trace(96fE+f(16R_{ijji}-8R_{amma}+13L_{aa}L_{bb}
$\par$+2L_{ab}L_{ab})I_V+f(96SL_{aa}+192S^2)
+f_{;m}(6L_{aa}I_V+96S)$\par$+24f_{;mm}I_V\}d\nu_{\partial M}$.
\smallskip\item $a_4^{\partial M}(f,D,\Cal{B})=(4\pi)^{-m/2}\frac1{360}
\int_{C_D}\trace\{f(-120E_{;m}+120EL_{aa})$\par$+f(-18R_{ijji;m}+20R_{ijji}L_{aa}
+4R_{amam}L_{bb}
-12R_{ambm}L_{ab}+4R_{abcb}L_{ac}$ 
\par
$+24L_{aa:bb}+\frac{40}{21}L_{aa}L_{bb}L_{cc}
-\frac{88}7L_{ab}L_{ab}L_{cc}+\frac{320}{21}L_{ab}L_{bc}L_{ac})I_V-180f_{;m}E
$\par$+f_{;m}(-30R_{ijji}
-\frac{180}{7}L_{aa}L_{bb}+\frac{60}7L_{ab}L_{ab})I_V
+24f_{;mm}L_{aa}I_V$
\par$-30f_{;iim}I_V\}d\nu_{\partial M}$
\par
$+(4\pi)^{-m/2}\frac1{360}\int_{C_N}\trace\{f(240E_{;m}+120EL_{aa})+f(42 R_{ijji;m}
+24L_{aa:bb}$
\par
$+20 R_{ijji}L_{aa}+4R_{amam}L_{bb}
-12R_{ambm}L_{ab}+4R_{abcb}L_{ac}+\frac{40}3L_{aa}L_{bb}L_{cc}$
\par
$+8L_{ab}L_{ab}L_{cc}+\frac{32}3L_{ab}L_{bc}L_{ac})I_V+f(720SE+120S R_{ijji}
+144SL_{aa}L_{bb}$
\par
$+48SL_{ab}L_{ab}+480S^2L_{aa}+480S^3+120S_{:aa})
+f_{;m}(180E+72SL_{aa}$\par$+240S^2)+f_{;m}(30R_{ijji}+12L_{aa}L_{bb}+12L_{ab}L_{ab}
)I_V+120f_{;mm}S$\par$+24f_{;mm}L_{aa}I_V+30f_{;iim}I_V\}d\nu_{\partial M}$.
\endroster\endproclaim

The main result of this paper is the
following result which extends Theorem \arefb\ to the time dependent setting:
\setref\arefc
\proclaim{\arefc\ Theorem}
\roster
\smallskip\item $a_0^M(f,\Cal{D})=a_0^M(f,D)$.
\smallskip\item $a_2^M(f,\Cal{D})=a_2^M(f,D)+(4\pi)^{-m/2}\frac16
\int_Mf\trace(\frac32\Cal{G}_{1,ii})d\nu_M$.
\smallskip\item $a_4^M(f,\Cal{D})=a_4^M(f,D)+(4\pi)^{-m/2}\frac1{360}\int_Mf\trace(
\frac{45}4\Cal{G}_{1,ii}\Cal{G}_{1,jj}+\frac{45}2\Cal{G}_{1,ij}\Cal{G}_{1,ij}$\par $
+60\Cal{G}_{2,ii}-180\Cal{E}_{1}
+15\Cal{G}_{1,ii}R_{jkkj}
  -30\Cal{G}_{1,ij}R_{ikkj}+90\Cal{G}_{1,ii}E
  +60\Cal{F}_{1,i;i}$\par $
 15\Cal{G}_{1,ii;jj}-30\Cal{G}_{1,ij;ij})d\nu_M$.
\smallskip\item $a_n^{\partial M}(f,\Cal{D},\Cal{B})=a_n^{\partial M}
(f,D,\Cal{B}_0)$ for $n\le2$.
\smallskip\item $a_3^{\partial M}(f,\Cal{D},\Cal{B})=
\textstyle a_3^{\partial M}(f,D,\Cal{B}_0)+(4\pi)^{(1-m)/2}\frac1{384}\int_{C_D}
    f\trace(-24\Cal{G}_{1,aa})d\nu_{\partial M}$
\par $+(4\pi)^{(1-m)/2}\frac1{384}\textstyle\int_{C_N}f\trace(24\Cal{G}_{1,aa})
   d\nu_{\partial M}$.
\par\item 
$a_4^{\partial M}(f,\Cal{D},\Cal{B})=
\textstyle a_4^{\partial M}(f,D,\Cal{B}_0)+(4\pi)^{-m/2}
\frac1{360}\int_{C_D}\trace\{f(30\Cal{G}_{1,aa}L_{bb}$\par$
-60\Cal{G}_{1,mm}L_{bb}+
  30\Cal{G}_{1,ab}L_{ab}+
30\Cal{G}_{1,mm;m}-30\Cal{G}_{1,aa;m}\textstyle+0\Cal{G}_{1,am;a}-30\Cal{F}_{1,m})
$\par
$+f_{;m}(-45\Cal{G}_{1,aa}+45\Cal{G}_{1,mm})\}d\nu_{\partial M}$
\par$\textstyle+(4\pi)^{-m/2}\frac1{360}\int_{C_N}\trace\{f(30\Cal{G}_{1,aa}L_{bb}
 +120\Cal{G}_{1,mm}L_{bb}
 -150\Cal{G}_{1,ab}L_{ab}$\par
$-60\Cal{G}_{1,mm;m}
+60\Cal{G}_{1,aa;m}+0\Cal{G}_{1,am;a}+150\Cal{F}_{1,m}
+180S\Cal{G}_{1,aa}$\par$-180S\Cal{G}_{1,mm}+360S_1+0T_{a:a})
+f_{;m}(45\Cal{G}_{1,aa}-45\Cal{G}_{1,mm})\}d\nu_{\partial M}$.
\endroster\endproclaim

Here is a brief outline to this paper. In \S\qctSB, we use invariance theory and dimensional analysis
 to study the general form of the invariants $a_n(f,\Cal{D},\Cal{B})$. We shall use
$\Cal{B}{}^-$ for Dirichlet and $\Cal{B}{}^+$ for Robin boundary conditions. We shall show,
for example, that there exist constants
$c_0$ and
$e_1^\pm$ so that:
$$\eqalign{
   &a_2^M(f,\Cal{D})=\textstyle a_2^M(f,D)
       +(4\pi)^{-m/2}\frac16\textstyle\int_Mf\trace(c_0\Cal{G}_{1,ii})d\nu_M
      \text{ and}\cr
   &a_3^{\partial M}(f,\Cal{D},\Cal{B})=a_3^{\partial M}(f,D,\Cal{B}_0)+\textstyle
    (4\pi)^{-(m-1)/2}\frac1{384}\int_{C_D}f\trace(e_1^-\Cal{G}_{1,aa})d\nu_{\partial M}\cr
    &\qquad\textstyle+(4\pi)^{-(m-1)/2}\frac1{384}
\int_{C_N}f\trace(e_1^+\Cal{G}_{1,aa})d\nu_{\partial M};}$$
we refer to Lemma \brefa\ for further details. The interior invariants will be described
by constants  $\{c_i\}_{i=0}^{10}$, the boundary invariants for Neumann boundary conditions
will be described by constants $\{e_i^+\}_{i=1}^{15}$, and the boundary invariants for
Dirichlet boundary conditions will be described by constants
$\{e_i^-\}_{i=1}^{11}$. We use the localizing function $f$ to decouple
the interior and the boundary integrals; with the exception of Lemma \brefc, there is no
interaction between the unknown constants $\{c_i\}$, $\{e_j^-\}$, and $\{e_k^+\}$. A
priori, those constants could depend on the dimension. In Lemma \brefb, we will use product
formulas to dimension shift and show the constants are dimension free. We complete the
proof of Theorem \arefc\ by evaluating these unknown constants; the values we shall derive
are summarized in Table
\brefba. 

We use various functorial properties to derive relations among these constants. For
example, in Lemma \brefc, we use the product 
formulas of Lemma \brefb\ to show that $c_5=10c_0$. The functorial
properties that these time dependent invariants satisfy and which are discussed 
in \S\qctSC-\S\qctSF\ are new and have not been used previously in other calculations of the
heat trace asymptotics. Thus we believe they are of interest in their own right. It is one
of the features of the functorial method that one has to work in great generality even if
one is only interested in special cases.  We found it necessary, for example, to consider
the very general time dependent boundary conditions of equation (\arefaC) to ensure that
the class of boundary conditions was invariant under the gauge and coordinate
transformations employed in \S\qctSD\ and \S\qctSE. We work with scalar operators as the
(possible) non-commutativity of the endomorphisms in the vector valued case plays no role
in the evaluation of $a_n$ for $n\le 4$.

We summarize the five functorial properties we shall use as follows. In \S\qctSB, we
consider a product manifold
$M=M_1\times M_2$ where $\partial M_2$ is empty, and an operator of the form 
$\Cal{D}=\Cal{D}_1\otimes
1+1\otimes\Cal{D}_2$. In Lemma \brefc, we show that
$$\eqalign{
   &a_n(f_1f_2,\Cal{D},\Cal{B})=\textstyle\sum_{p+q=n}
   a_p(f_1,\Cal{D}_1,\Cal{B})a_q(f_2,\Cal{D}_2).}$$
In \S\qctSC, we rescale the time parameter $t$. Let
$D$ and $\Cal{B}$ be static operators. Let
$\Cal{D}:=(1+2\alpha t+3\beta t^2)D$. In Lemma \crefa, we show that: 
$$\eqalign{
   &\textstyle a_2(f,\Cal{D},\Cal{B})=a_2(f,D,\Cal{B})-{m\over2}\alpha a_0(f,D,\Cal{B})\cr
   &\textstyle a_3(f,\Cal{D},\Cal{B})=a_3(f,D,\Cal{B})-
   \textstyle\frac{m-1}2\alpha a_1(f,D,\Cal{B})\cr
   &\textstyle a_4(f,\Cal{D},\Cal{B})=a_4(f,D,\Cal{B})
    -{m-2\over2}\alpha a_2(f,D,\Cal{B})
    +({m(m+2)\over8}\alpha^2-{m\over2}\beta)a_0(f,D,\Cal{B}).}$$
In \S\qctSD, we make a time dependent gauge transformation. We assume $D$ and $\Cal{B}$
are static. Let $\Cal{D}_\varrho:=e^{-t\varrho\Psi}De^{t\varrho\Psi}+\varrho\Psi$. We also
gauge transform the boundary condition $\Cal{B}$ to define $\Cal{B}_\varrho$. In Lemma
\drefa, we show that :
$$\birdy{a_n(f,\Cal{D}_\varrho,\Cal{B}_\varrho)}
     =-a_{n-2}(f\Psi,D,\Cal{B}).$$
In \S\qctSE, we make a time dependent coordinate transformation. Let $\Delta$ be the
scalar Laplacian and let $\Cal{B}$ be static.
Let $\Phi_\varrho:(t,x_1,x_2)\rightarrow(t,x_1+t\varrho\Xi,x_2)$ where $\varrho$ is an
auxiliary parameter. We set
$\Cal{D}_\varrho:=\Phi_{\varrho}^*(\partial_t+\Delta)
-\partial_t$ and $\Cal{B}_\varrho:=\Phi_{\varrho}^*(\Cal{B})$. Let
$d\nu_M:=gdx^1dx^2$. In Lemma \erefa, we show that:
$$\birdy{a_n(f,\Cal{D}_\varrho,\Cal{B}_\varrho)}
    =-\textstyle\frac12a_{n-2}(g^{-1}\partial_1(gf\Xi),\Delta,\Cal{B}).$$
In \S\qctSF, we assume given a second order operator $Q$ which commutes with a static
operator $D$ of Laplace type. We define $D_\varrho:=D+\varrho Q$ and define a suitable
boundary condition $\Cal{B}_\varrho$. We also define
$\Cal{D}_\varrho:=D+2t\varrho Q$ and show
$$\birdy{a_n(f,\Cal{D}_\varrho,\Cal{B})}=\birdy{a_{n-2}(f,D_\varrho,\Cal{B}_\varrho)}.$$

In each section, we use the relevant functorial properties to derive relations among the
unknown coefficients; these relations are contained in Lemmas \brefc, \crefb, \drefb, and
\erefb. These relations suffice to determine the unknown coefficients and thereby complete
the proof of Theorem \arefc. As the computations are somewhat long and technical, we have
derived more equations than are needed as a consistency check; this is typical in such
computations. 
\sethead\qctSB\head\S\qctSB\ Invariance Theory, dimensional analysis, and dimension shifting\endhead

We begin the proof of Theorem \arefc\ by establishing the 
general form of the invariants
$a_n^M$ and $a_n^{\partial M}$ for $n\le4$. Let $(D,\Cal{B}_0)$ be the
static operator and boundary condition determined by $(\Cal{D},\Cal{B})$.
\setref\brefa
\proclaim{\brefa\ Lemma} There exist constants so that
\roster
\smallskip\item $a_0^M(f,\Cal{D})=a_0^M(f,D)$ and 
$a_i^{\partial M}(f,\Cal{D},\Cal{B})=a_i^{\partial M}(f,D,\Cal{B}_0)$ for $i\le 2$.
\smallskip\item $a_2^M(f,\Cal{D})=a_2^M(f,D)+(4\pi)^{-m/2}
\textstyle{1\over6}\int_Mf\trace\{c_0\Cal{G}_{1,ii}\}d\nu_M$.
\smallskip\item $a_4^M(f,\Cal{D})=a_4^M(f,D)+(4\pi)^{-m/2}
\textstyle{1\over360}\int_Mf\trace\{
   c_1\Cal{G}_{1,ii}\Cal{G}_{1,jj}+c_2\Cal{G}_{1,ij}\Cal{G}_{1,ij}$
\par
  $+c_3\Cal{G}_{2,ii}
  +c_4\Cal{E}_{1}+c_5\Cal{G}_{1,ii}R_{jkkj}
  +c_6\Cal{G}_{1,ij}R_{ikkj}+c_7\Cal{G}_{1,ii}E
  +c_8\Cal{F}_{1,i;i}$\par$
 +c_9\Cal{G}_{1,ii;jj}+c_{10}\Cal{G}_{1,ij;ij}
  \}d\nu_M$.
\smallskip\item $a_3^{\partial M}(f,\Cal{D},\Cal{B})=
\textstyle a_3^{\partial M}(f,D,\Cal{B}_0)+(4\pi)^{(1-m)/2}\frac1{384}
\int_{C_D}f\trace(e_1^-\Cal{G}_{1,aa}$
\par$+e_2^-\Cal{G}_{1,mm})d\nu_{\partial M}
+(4\pi)^{(1-m)/2}\frac1{384}\textstyle\int_{C_N}f\trace(e_1^+\Cal{G}_{1,aa}
+e_2^+\Cal{G}_{1,mm})d\nu_{\partial M}$.
\smallskip\item 
$a_4^{\partial M}(f,\Cal{D},\Cal{B})=\textstyle a_4(f,D,\Cal{B}_0)+(4\pi)^{-m/2}
\frac1{360}\int_{C_D}\trace\{f(e_3^-\Cal{G}_{1,aa}L_{bb}$\par$
+e_4^-\Cal{G}_{1,mm}L_{bb}+
  e_5^-\Cal{G}_{1,ab}L_{ab}+
e_6^-\Cal{G}_{1,mm;m}+e_7^-\Cal{G}_{1,aa;m}$\par
$\textstyle+e_8^-\Cal{G}_{1,am;a}+e_9^-\Cal{F}_{1,m})
+f_{;m}(e_{10}^-\Cal{G}_{1,aa}+e_{11}^-\Cal{G}_{1,mm})\}d\nu_{\partial M}$
\par$\textstyle+(4\pi)^{-m/2}\frac1{360}\int_{C_N}
\trace\{f(e_3^+\Cal{G}_{1,aa}L_{bb}+e_4^+\Cal{G}_{1,mm}L_{bb}+
  e_5^+\Cal{G}_{1,ab}L_{ab}$\par
$+e_6^+\Cal{G}_{1,mm;m}
+e_7^+\Cal{G}_{1,aa;m}+e_8^+\Cal{G}_{1,am;a}+e_9^+\Cal{F}_{1,m}
+e_{12}^+S\Cal{G}_{1,aa}$\par$
+e_{13}^+S\Cal{G}_{1,mm}+e_{14}^+S_1+e_{15}^+T_{a:a})
+f_{;m}(e_{10}^+\Cal{G}_{1,aa}+e_{11}^+\Cal{G}_{1,mm})\}d\nu_{\partial M}$.
\endroster
\endproclaim

\demo{Proof} We use dimensional analysis - this involves studying the behavior of
these invariants under rescaling and is described in \cite{\refGilkey} in the static
setting. We assign weight 2 to $R$, $\Omega$, $E$ and $T_a$ and 
weight 3 to $S_1$. We assign weight
$1$ to
$S$ and $L_{ab}$. We increase the weight by
$1$ for each explicit covariant derivative which appears. Thus, for example, the
terms $E_{;kk}$, $\Omega_{ij}\Omega_{ij}$, and $R_{ijkl}R_{ijkl}$ are all of degree
$4$.  The integrands appearing in $a_n^M$ and $a_n^{\partial M}$ are weighted
homogeneous of degree $n$ and $n-1$.  The structure groups are $O(m)$ and $O(m-1)$
respectively. H. Weyl's Theorem \cite{\refWeyl} shows that all orthogonal invariants
are given by contractions of indices. The assertions of the Lemma now follow by
writing down a spanning set for the space of invariants. We remark that since
$\Cal{G}_{1,ij}=\Cal{G}_{1,ji}$, the invariant
$\Cal{G}_{1,ij}\Omega_{ij}$ does not appear. \qed\enddemo
We will complete the proof of Theorem \arefc\ by evaluating the unknown
coefficients of Lemma \brefa. The remainder of this paper is devoted to deriving the
values in the following table:
\setref\brefba
\medbreak\centerline{\bf Table \brefba}
\medbreak\centerline{\vrule\vbox{\offinterlineskip\hrule\halign{
\ #\ \hfil&\vrule height12pt
\ #\ \hfil&\vrule height12pt\ #\ \hfil&\vrule height12pt\ #\ \hfil&\vrule height12pt\ #\ \hfil&\vrule
height12pt\ #\ \hfil\cr
  $c_{0{\phantom{;}}}=\frac32$&$c_1=\frac{45}4$&$c_2=\frac{45}2$&$c_3=60$&$c_4=-180$&$c_{5}=15$\cr\noalign{\hrule}
  $c_{6{\phantom{;}}}=-30$&$c_7=90$&$c_8=60$&$c_9=15$&$c_{10}=-30$&\cr\noalign{\hrule}
  $e_{1{\phantom{;}}}^-=-24$&$e_2^-=0$&$e_3^-=30$&$e_4^-=-60$&$e_5^-=30$&$e_6^-=30$\cr\noalign{\hrule}
  $e_{7{\phantom{;}}}^-=-30$&$e_8^-=0$&$e_9^-=-30$&$e_{10}^-=-45$&$e_{11{\phantom{;}}}^-=45$&\cr\noalign{\hrule}
  $e_{1{\phantom{;}}}^+=24$&$e_2^+=0$&$e_3^+=30$&$e_4^+=120$&$e_5^+=-150$&$e_6^+=-60$\cr\noalign{\hrule}
  $e_{7{\phantom{;}}}^+=60$&$e_8^+=0$&$e_9^+=150$&$e_{10}^+=45$&$e_{11}^+=-45$&$e_{12}^+=180$\cr\noalign{\hrule}
  $e_{13}^+=-180$&$e_{14}^+=360$&$e_{15}^+=0$&&&\cr
}\hrule}\vrule}
\medbreak The (possible) non-commutativity of the endomorphisms in the vector valued
case plays no role in the invariants of Lemma \brefa. We therefore suppose $V$ to be
the trivial bundle hence forth and omit the trace from our formulas to simplify the
notation as we will be dealing with scalar operators on $C^\infty(M)$. We also set
$e_i^-=0$ for $i\ge12$ to have a common formalism; these constants describe
invariants which involve $S$, $S_1$, and $T_a$ and which are therefore not relevant
for Dirichlet boundary conditions.

A-priori, the constants $c_i$ and $e_i^\pm$ might depend upon the
dimension. Fortunately, this turns out not to be the case; the dependence upon the
dimension is contained in the multiplicative normalizing factors of $(4\pi)^*$. Let
$\Cal{D}_i$ be smooth time dependent families of operators of Laplace type over
manifolds $M_i$ for
$i=1,2$. We suppose $M_2$ is closed. Let $M:=M_1\times M_2$, let
$\Cal{D}:=\Cal{D}_1+\Cal{D}_2$, and let the boundary condition for $M$ be induced
from the corresponding boundary condition for $M_1$.

\setref\brefb
\proclaim{\brefb\ Lemma} Adopt the notation established above.\roster
\smallskip\item 
$a_n^M(f_1 f_2,\Cal{D})=\textstyle\sum_{p+q=n}
    a_p^{M_1}(f_1,{\Cal D}_1)a_q^{M_2}(f_2,{\Cal D}_2)$
\smallskip\item
$a_n^{\partial M}(f_1 f_2,\Cal{D},\Cal{B})=
    \textstyle\sum_{p+q=n}a_p^{\partial M_1}(f_1,{\Cal D}_1,\Cal{B})
    a_q^{M_2}(f_2,{\Cal D}_2)$.
\smallskip\item The constants of Lemma 
    \brefa\ do not depend upon the dimension $m$.
\endroster\endproclaim

\demo{Proof} We use equation (\arefad) to check that
$u_{\phi_1\cdot\phi_2}=u_{\phi_1}\cdot u_{\phi_2}$. This shows the kernel function on
$M$ is the product of the corresponding kernel functions on $M_1$ and on $M_2$;
assertions (1) and (2) now follow. Let $(M,\Cal{D}_M,\Cal{B})$ be given. Let $S^1$
be the unit circle with the usual flat metric and usual periodic parameter $\theta$.
Let
$D_S=-\partial_\theta^2$ on the trivial line bundle. Let
$\Cal{D}_{M\times S^1}=\Cal{D}_M+D_S$. Then
$a_p(\theta,D_S)=0$ for $p>0$ and
$a_0(\theta,D_S)=(4\pi)^{-1/2}$; see \cite{\refGilkey} for details. 
Thus $p=n$ and $q=0$ in assertions (1) and (2) so
$a_n(f_1,\Cal{D}_{M\times S^1})=(4\pi)^{-1/2}a_n(f_1,\Cal{D}_M,\Cal{B})$.
It now follows that $c_i(m+1)=c_i(m)$ and $e_i^\pm(m+1)=e_i^\pm(m)$.
\qed\enddemo

\medbreak We use the product formulas of Lemma \brefb\ to prove the following Lemma:
\setref\brefc
\proclaim{\brefc\ Lemma} We have $c_1=5c_0^2$, $c_5=10c_0$, $c_7=60c_0$,
$e_1^-=-16c_0$, $e_3^-=20c_0$, $e_{10}^-=-30c_0$, $e_1^+=16c_0$, $e_3^+=20c_0$,
$e_{10}^+=30c_0$, and $e_{12}^+=120c_0$.
\endproclaim

\demo{Proof} We apply Lemma \brefb\ and study the cross terms
arising  in 
$a_{p+q}(f_1f_2,\Cal{D},\Cal{B})$ from
$a_p(f_1,\Cal{D}_1,\Cal{B}_1)a_q(f_2,\Cal{D}_2)$. We let indices
$r$ and $s$ index $M_1$ and indices $u$ and $v$ index $M_2$. We use Theorem
\arefb\ and equate coefficients of
suitable expressions to derive the following systems of equations from which the
Lemma will follow:\medbreak\centerline{\vrule\vbox{\offinterlineskip\hrule\halign{
\phantom{$\frac1{6_{\vrule height 6pt}}$} #\ \hfil&\vrule height14pt
\ #\ \hfil&\vrule height12pt\ #\ \hfil&\vrule height12pt\ #\ \hfil&\vrule height12pt\ #\ \hfil&\vrule
height12pt\ #\ \hfil\cr
$2c_1=360(\frac16c_0)(\frac16c_0)$\quad
                 \hfill [$f_1f_2\Cal{G}_{1,rr}\Cal{G}_{1,uu}$]
&$c_5=360(\frac16)(\frac16c_0)$\hfill [$f_1f_2R_{rssr}\Cal{G}_{1,uu}$]\cr
\noalign{\hrule}
$c_7=360(\frac16c_0)$\hfill [$f_1f_2E_1\Cal{G}_{1,uu}$]
&$e_1^\pm=384(\pm\frac14)(\frac16c_0)$
     \hfill [$f_1f_2\Cal{G}_{1,uu}$]\cr\noalign{\hrule}
$e_3^\pm=360(\frac13)(\frac16c_0)$
    \hfill [$f_1f_2L_{rr}\Cal{G}_{1,uu}$]
&$e_{10}^\pm=360(\pm\frac12)(\frac16c_0)$\quad
     \hfill [$f_{1;m}f_2\Cal{G}_{1,uu}$]\cr\noalign{\hrule}
$e_{12}^+=360(2)(\frac16c_0)$
     \hfill [$fS\Cal{G}_{1,uu}$]\ \qed&
\cr
}\hrule}\vrule}\enddemo

\sethead\qctSC\head\S\qctSC\ Rescaling the time parameter\endhead

Let $D$ and $\Cal{B}$ be static. Let $\alpha,\beta\in\Bbb{R}$. We define a time dependent
family of operators of Laplace type by setting: $\Cal{D}:=(1+2\alpha t+3\beta t^2)D$.

\setref\crefa
\proclaim{\crefa\ Lemma}
\roster
\smallskip\item $a_2(f,\Cal{D},\Cal{B})=a_2(f,D,\Cal{B})-{m\over2}\alpha
a_0(f,D,\Cal{B})$.
\smallskip\item $a_3(f,\Cal{D},\Cal{B})=a_3(f,D,\Cal{B})-
   \textstyle\frac{m-1}2\alpha a_1(f,D,
\Cal{B})$.
\smallskip\item $a_4(f,\Cal{D},\Cal{B})=a_4(f,D,\Cal{B})
    -{m-2\over2}\alpha a_2(f,D,\Cal{B})
    +({m(m+2)\over8}\alpha^2-{m\over2}\beta)a_0(f,D,\Cal{B})$.
\endroster\endproclaim
\demo{Proof}
Let $u_0=e^{-tD_{\Cal{B}}}\phi$ and let
$u(t,x):=u_0(t+\alpha t^2+\beta t^3,x)$. Then:
$$\eqalign{
   &\Cal{D}u(t,x)=(1+2\alpha t+3\beta t^2)(Du_0)(t+\alpha t^2+\beta t^3,x)\cr
   &\partial_tu(t,x)=(1+2\alpha t+3\beta t^2)(\partial_tu_0)(t+\alpha t^2+\beta t^3,x).}$$
This shows that $(\partial_t+\Cal{D})u=0$. Since $u(0,x)=u_0(0,x)=\phi(x)$ and
$\Cal{B}u=0$, the relations of equation
(\arefad) are satisfied so
$$K(t,x,\bar x,\Cal{D},\Cal{B})=K(t+\alpha t^2+\beta t^3,x,\bar
x,D,\Cal{B}).$$
The Lemma will
then follow from the expansions:
$$\eqalign{
    &a(f,\Cal{D},\Cal{B})(t)\sim\textstyle\sum_nt^{-m/2}
    (1+\alpha t+\beta t^2)^{(n-m)/2}a_n(f,D,\Cal{B})t^{n/2}\cr
    &(1+\alpha t+\beta t^2)^j\sim
     \textstyle1+\alpha jt+({j(j-1)\over2}\alpha^2+j\beta)t^2+O(t^3)
    \ \qed}$$\enddemo

We apply Theorem \arefb\ and Lemma \crefa\ to derive the following relationships:
\setref\crefb
\proclaim{\crefb\ Lemma}
\roster
\smallskip\item $c_0=\textstyle{3\over2}$, $c_1={45\over4}$, $c_2={45\over2}$, 
  $c_3=60$, $c_4=-180$, $c_5=15$, $c_6=-30$, $c_7=90$.
\smallskip\item $e_1^\pm=\pm24$, $e_2^\pm=0$, $e_3^\pm=30$, $e_4^\pm+e_5^\pm=-30$, 
$e_{10}^\pm=\pm45$,
$e_{11}^\pm=\mp45$.
\smallskip\item $e_{12}^+=180$,
$e_{13}^+=-180$.
\endroster\endproclaim

\demo{Proof}
We have $\Cal{G}_{1,ij}=-2\alpha g_{ij}$, $\Cal{F}_{1,i}=0$, 
$\Cal{G}_{2,ij}=-3\beta g_{ij}$,
and $\Cal{E}_1=-2\alpha E$. Thus $\Cal{G}_{1,ii;jj}=0$, $\Cal{G}_{1,ij;ij}=0$, and
$\Cal{F}_{1,i;i}=0$. We equate coefficients of suitable expressions in Lemma \crefa\ to
derive the following systems of equations from which the Lemma will
follow. Note that since $m$ is
arbitrary, equations involving this
parameter can give rise to more than one
relation. 
\medbreak\centerline{\vrule\vbox{\offinterlineskip\hrule\halign{
\phantom{$\frac1{6_{\vrule height 6pt}}$} #\ \hfil&\vrule height14pt\ \hfill#\ 
\cr $-2mc_0=-6{m\over2}$ 
                 & [$\alpha f$] in $a_2^M$
\cr\noalign{\hrule} $4(m^2c_1+mc_2)=360{m(m+2)\over8}$
                 & [$\alpha^2f$] in $a_4^M$
\cr\noalign{\hrule} $-3mc_3=-360{m\over2}$ 
                 & [$\beta f$] in $a_4^M$
\cr\noalign{\hrule} $-2(c_4+mc_7)=-360{m-2\over12}6$
                 & [$\alpha fE$] in $a_4^M$
\cr\noalign{\hrule} $-2(mc_5+c_6)=-360{m-2\over12}$
                 & \qquad[$\alpha fR_{ijji}$] in $a_4^M$
\cr\noalign{\hrule} $-2\{(m-1)e_1^\pm+e_2^\pm\}=-384(\frac{m-1}2)(\pm\frac14)$&[$\alpha f$]
    in $a_3^{\partial M}$
\cr\noalign{\hrule}$-2\{(m-1)e_3^\pm+e_4^\pm+e_5^\pm\}=-360(\frac{m-2}2)(\frac13)$&[$\alpha fL_{aa}$]
  in $a_4^{\partial M}$
\cr\noalign{\hrule}$-2\{(m-1)e_{10}^\pm+e_{11}^\pm\}=-360(\frac{m-2}2)(\pm\frac12)$&[$\alpha f_{;m}$]
in $a_4^{\partial M}$
\cr\noalign{\hrule}$-2\{(m-1)e_{12}^++e_{13}^+\}=-360(\frac{m-2}2)(2)$.&[$\alpha fS$] in $a_4^{\partial
M}$\qed
\cr
}\hrule}\vrule}
\enddemo

\sethead\qctSD\head\S\qctSD\ Time dependent gauge transformations\endhead

Let $\Cal{D}_\varrho:=e^{-t\varrho\Psi}De^{t\varrho\Psi}+\varrho\Psi$. If 
$\Cal{B}u=u_{;m}+Su$ is the
Robin boundary operator, we gauge transform the boundary condition
to define $\Cal{B}_\varrho:=\nabla_m+S+tS_1$ with $S_1=\varrho\Psi_{;m}$;
the Dirichlet boundary operator is unchanged.

\setref\drefa
\proclaim{\drefa\ Lemma} We have $\birdy{a_n(f,\Cal{D}_\varrho,\Cal{B}_\varrho)}
     =-a_{n-2}(f\Psi,D,\Cal{B})$.
\endproclaim

\demo{Proof} Let $u_0:=e^{-tD_{\Cal B}}\phi$ and let
$u:=e^{-t\varrho\Psi}u_0$. We show $u$ satisfies the relations of (\arefad) by computing:
$$\eqalign{
&\partial_t u(t,x)=e^{-t\varrho\Psi}(\partial_t-\varrho\Psi)u_0,\ 
\Cal{D}_\varrho u(t,x)=e^{-t\varrho\Psi}(D+\varrho\Psi)u_0,\cr
&(\partial_t+\Cal{D}_\varrho )u=e^{-t\varrho\Psi}(\partial_t+D)u_0=0,
\text{ and }u(0,x)=u_0(x)=\phi(x).
}$$
Dirichlet boundary conditions are preserved. With Robin boundary conditions,
$$ u_{;m}+S u+tS_1 u=e^{-t\varrho\Psi}(u_{0;m}-t\varrho\Psi_{;m}u_0+Su_0+
t \varrho\Psi_{;m}u_0)=0.$$
Thus $K(\cdot,\Cal{D}_\varrho,\Cal{B}_\varrho)=e^{-t\varrho\Psi}K(\cdot,D,\Cal{B})$. The
Lemma now follows. \qed
\enddemo

We use Lemma \drefa\ to obtain some additional relationships:
\setref\drefb
\proclaim{\drefb\ Lemma} We have $c_8=60$, $e_9^-=-30$, and
$e_{14}^+-2e_9^+=60$.\endproclaim

\demo{Proof} Let $\Psi$ vanish on $\partial M$. We apply Lemma \drefa\ with $M=[0,1]$ and
$D=-\partial_\theta^2$. We work modulo terms which are $O(\varrho^2)$ and compute:
$$\eqalign{
&D_\varrho\equiv D+\varrho\Psi-2t\varrho\Psi_{;\theta}\partial_\theta
  -t\varrho\Psi_{;\theta\theta},\cr
&\Cal{B}_\varrho^+\equiv\nabla_m+S+t\varrho\Psi_{;m},\ S_1\equiv\varrho\Psi_{;\theta},\cr 
&E\equiv-\varrho\Psi,\
\Cal{F}_{1,m}\equiv-2\varrho\Psi_{;\theta},\ 
 \Cal{E}_{1}\equiv-\varrho\Psi_{;\theta\theta}.}$$
We study $\birdy{a_4^M}$ and $\birdy{a_4^{\partial M}}$:
\medbreak\centerline{\vrule\vbox{\offinterlineskip\hrule\halign{
\phantom{$\frac1{6_{\vrule height 6pt}}$} #\ \hfil&\vrule height14pt\ #\hfil\ 
\cr
 $\birdy{60E_{;ii}}\equiv-60\Psi_{;\theta\theta}$
&$\birdy{-180\Cal{E}_1}\equiv180\Psi_{;\theta\theta}$\cr\noalign{\hrule}
 $\birdy{c_8\Cal{F}_{1,i;i}}\equiv-2c_8\Psi_{;\theta\theta}$
&$\birdy{e_{14}^+S_1}\equiv e_{14}^+\Psi_{;\theta}$ \cr\noalign{\hrule}
 $\birdy{(-120^-,240^+)E_{;m}}\equiv(120^-,-240^+)\Psi_{;\theta}$
&$\birdy{e_9^\pm\Cal{F}_{1,m}}\equiv-2e_9^\pm\Psi_{;\theta}$
\cr
}\hrule}\vrule}
\medbreak\noindent Here the notation $(-120^-,240^+)$ indicates that the coefficient for
Dirichlet $\Cal{B}^-$ and Neumann $\Cal{B}^+$ boundary conditions is $-120$ and
$240$. As
$-a_2^M(f\Psi,D)=0$ and $-a_2^{\partial
M}(f\Psi,D,\Cal{B}^\pm)=-\frac1{360}(4\pi)^{-1/2}\textstyle\int_{\partial M}
    \pm180(f\Psi)_{;m}$, we use Lemma \drefa\ to derive the following equations from which
the Lemma will follow:
$$\eqalign{0&=-60+180-2c_8,\cr-180&=-2e_9^++e_{14}^+-240,\cr 
 180&=120-2e_9^-.\ \qed}$$
\enddemo

\sethead\qctSE\head\S\qctSE\ Time dependent coordinate transformations\endhead
In this section, we study time dependent coordinate transformations and make a coordinate
transformation that mixes up the spatial and the temporal coordinates. This technique was
also used in \cite{\refGilkeya} to study the heat content asymptotics. We work in a very
specific context but note the Lemma holds true in much greater generality. Let
$M:=S^1\times[0,1]$ with
$ds^2=e^{2\psi_1}dx_1^2+e^{2\psi_2}dx_2^2$. Let
$d\nu_M:=gdx^1dx^2$. Let
$\Xi\in C^\infty(M)$ have compact support near some point $P\in M$.
Let $\Delta$ be the scalar Laplacian and let $\Cal{B}$ be a static boundary condition.
Define:
$$\eqalign{
  &\Phi_\varrho(t,x_1,x_2):=(t,x_1+t\varrho\Xi,x_2),\cr
  &\Cal{D}_\varrho:=\Phi_\varrho^*(\partial_t+\Delta)-\partial_t,\text{ and }
   \Cal{B}_\varrho:=\Phi_\varrho^*(\Cal{B}).}$$
\setref\erefa
\proclaim{\erefa\ Lemma} We have
    $\frac\partial{\partial\varrho}a_n(f,\Cal{D}_\varrho,\Cal{B}_\varrho)|_{\varrho=0}=
      -\frac12a_{n-2}(g^{-1}\partial_1(gf\Xi),\Delta,\Cal{B})$.
\endproclaim
\demo{Proof} Let $u(t,x_1,x_2):=\{\Phi_\varrho^*(e^{-t\Delta_{\Cal{B}}}\phi)\}(x_1,x_2)$. By
naturality, $u$ satisfies the relations of (\arefad). As the static operator determined by
$\Cal{D}_\varrho$ is $\Delta+$ lower order terms,
$d\nu_M$ is independent of $\varrho$. Thus
$$K(t,x_1,x_2,\bar x_1,\bar x_2,\Cal{D}_\varrho,\Cal{B}_\varrho)
   =K(t,x_1+\varrho t\Xi(x_1,x_2),x_2,\bar x_1,\bar x_2,\Delta,\Cal{B}).$$
We set $x_1=\bar x_1$ and $x_2=\bar x_2$. We work modulo terms which are $O(\varrho^2)$
and expand in a Taylor series to compute:
$$\eqalign{
    &a(f,\Cal{D}_\varrho,\Cal{B}_\varrho)(t)=
    \textstyle\int_M f(x_1,x_2)K(t,x_1,x_2,x_1,x_2,\Cal{D}_\varrho,\Cal{B}_\varrho)d\nu_M\cr
   &\quad=\textstyle\int_M f(x_1,x_2)K(t,x_1+\varrho t\Xi,x_2,x_1,x_2,\Delta,\Cal{B})
     gdx_1dx_2\cr
   &\quad\equiv\textstyle\int_M\{f(x_1,x_2)K(t,x_1,x_2,x_1,x_2,\Delta,\Cal{B})\cr
    &\qquad+t\varrho f\Xi\partial_1K(t,x_1,x_2,y_1,x_2,\Delta,\Cal{B})|_{x_1=y_1}\}
   gdx_1dx_2.}$$ 
As $\Delta_{\Cal B}$ is self adjoint, the heat
kernel is symmetric. Thus we have:
$$\eqalign{&a(f,\Cal{D}_\varrho,\Cal{B}_\varrho)(t)
   \equiv\textstyle\int_M\{f(x_1,x_2)K(t,x_1,x_2,x_1,x_2,\Delta,\Cal{B})\cr
    &\textstyle\qquad+\frac12t\varrho f\Xi\partial_1K(t,x_1,x_2,x_1,x_2,\Delta,\Cal{B})\}
   gdx_1dx_2\cr
   &\quad\equiv\textstyle\int_M\{f(x_1,x_2)K(t,x_1,x_2,x_1,x_2,\Delta,\Cal{B})\cr
    &\qquad-\textstyle\frac12t\varrho g^{-1}\partial_1(gf\Xi)
    K(t,x_1,x_2,x_1,x_2,\Delta,\Cal{B})\}
      d\nu_M\cr
  &\quad\equiv a(f,\Delta,\Cal{B})
       (t)-\textstyle\frac12t\varrho a
(g^{-1}\partial_1(gf\Xi),\Delta,\Cal{B})(t).\ \qed}$$\enddemo
We use Lemma \erefa\ to complete the proof of Theorem \arefc\ by completing the
calculation of the coefficients $c_i$ and $e_i^\pm$.
\setref\erefb
\proclaim{\erefb\ Lemma}\roster
\smallskip\item $c_9=15$ and $c_{10}=-30$.
\smallskip\item $e_4^-=-60$, $e_5^-=30$, $e_6^-=30$, $e_7^-=-30$, and $e_8^-=0$.
\smallskip\item $e_4^+=120$, $e_5^+=-150$, $e_6^+=-60$, $e_7^+=60$, $e_8^+=0$, $e_9^+=150$, $e_{14}^+=360$, and
   $e_{15}^+=0$.
\endroster\endproclaim

\demo{Proof} We introduce an auxiliary parameter $\varepsilon$ and work modulo terms which
are $O(\varepsilon^2)+O(\varrho^2)$. Let
$$ds^2:=e^{2\varepsilon\psi_1}dx_1^2+e^{2\varepsilon\psi_2}dx_2^2.$$
The Laplacian $\Delta=-g^{-1}\partial_igg^{ij}\partial_j$ can then be expressed in the form
$$\Delta\equiv-\{e^{-2\varepsilon\psi_1}\partial_1^2
   +e^{-2\varepsilon\psi_2}\partial_2^2+\varepsilon(\psi_{2/1}
   -\psi_{1/1})\partial_1+\varepsilon(\psi_{1/2}
   -\psi_{2/2})\partial_2\}.$$
Let $\Phi_\varrho(t,x_1,x_2)=(t,x_1+\varrho t \Xi,x_2)$. Let $\Xi_{/i}=\partial_i\Xi$ etc. 
As $\Phi_\varrho$ is a diffeomorphism, we can pull back both differential forms and
differential operators. We compute:
$$\eqalign{
   &\Phi_\varrho^*(\partial_1)\equiv\partial_1-t\varrho \Xi_{/1}\partial_1,\ 
    \Phi_\varrho^*(\partial_2)\equiv\partial_2-t\varrho \Xi_{/2}\partial_1,
    \Phi_\varrho^*(\partial_t)\equiv\partial_t-\varrho\Xi\partial_1.}$$
The operator $\Cal{D}_\varrho:=\Phi_\varrho^*(\partial_t+\Delta)-\partial_t$
is given by:
$$\eqalign{\Cal{D}_\varrho\equiv&\Delta+t\varrho\{
    e^{-2\epsilon\psi_1}[2\Xi_{/1}\partial_1^2
   +\Xi_{/11}\partial_1
   ] +e^{-2\epsilon \psi_2} [2\Xi_{/2}\partial_1\partial_2
        +\Xi_{/22}\partial_1]\}\cr
   & +t\varrho\varepsilon\{2\psi_{1/1}\Xi\partial_1^2
   +2\psi_{2/1}\Xi\partial_2^2
    +\Xi_{/1}(\psi_{2/1}-\psi_{1/1})\partial_1\cr
    &-\Xi(\psi_{2/11}-\psi_{1/11})\partial_1
    +\Xi_{/2}(\psi_{1/2}-\psi_{2/2})\partial_1\cr
    &-\Xi(\psi_{1/12}-\psi_{2/12})\partial_2\}.}$$
 The tensors  $E$, $\Cal{G}$, and $\Cal{E}_1$ are therefore given by:
\medbreak\centerline{\vrule\vbox{\offinterlineskip\hrule\halign{
\phantom{$\frac1{6_{\vrule height 6pt}}$} #\ \hfil&\vrule height14pt\ #\hfill\ 
\cr $\textstyle\Cal{D}_0=\Delta-\varrho\Xi\partial_1$& 
    $\omega_1^{\Cal{D}}\equiv\frac12e^{2\varepsilon\psi_1}\varrho\Xi$\cr\noalign{\hrule}
    $\textstyle\Cal{G}_{1,}{}^{11}\equiv
     e^{-2\varepsilon\psi_1}2\varrho\Xi_{/1}+2\varepsilon\psi_{1/1}\varrho\Xi$&
    $\omega_2^{\Cal{D}}\equiv0$\cr\noalign{\hrule}
    $\Cal{G}_{1,}{}^{22}\equiv 2\varepsilon\psi_{2/1}\varrho\Xi$&
    $\Cal{G}_{1,}{}^{12}\equiv e^{-2\varepsilon\psi_2}\varrho\Xi_{/2}$
     \cr\noalign{\hrule}
    $\textstyle E\equiv-\frac12\varrho\Xi_{/1}
     -\frac12\varepsilon(\psi_{1/1}+\psi_{2/1})\varrho\Xi$&
     $\Cal{E}_1\equiv0$\cr}\hrule}\vrule}

\medbreak\noindent To compute $\Cal{F}$, we must express partial differentiation in
terms of covariant differentiation. Since $\omega$ is linear in $\varrho$, it plays no role.
The Christoffel symbols of the metric, however, play a crucial role. We compute:
$$\eqalign{
   &\Cal{G}_{1,}{}^{11}f_{;11}
        \equiv(\Cal{G}_{1,}{}^{11}\partial_1^2-2\varrho\Xi_{/1}\varepsilon\psi_{1/1}\partial_1
   +2\varrho\Xi_{/1}\varepsilon\psi_{1/2}\partial_2)f\cr
   &2\Cal{G}_{1,}{}^{12}f_{;12}
       \equiv (2\Cal{G}_{1,}{}^{12}\partial_1\partial_2
   -2\varrho\Xi_{/2}\varepsilon\psi_{1/2}\partial_1
   -2\varrho\Xi_{/2}\varepsilon\psi_{2/1}\partial_2) f\cr
   &\Cal{G}_{1,}{}^{22}f_{;22}\equiv\Cal{G}_{1,}{}^{22}\partial_2^2f\cr}$$
We use this computation to determine the tensor $\Cal{F}_1$:
   \medbreak\centerline{\vbox{\offinterlineskip\halign{
\phantom{${}_{\vrule height 8pt}^{\vrule height 8pt}$} #\ \hfil&\ #\hfill\ 
\cr $\Cal{F}_{1,}{}^1\equiv
       \varrho(e^{-2\varepsilon \psi_1}\Xi_{/11}+
     e^{-2\varepsilon \psi_2}\Xi_{/22})$\cr\qquad\qquad$
       +\varepsilon\varrho\{(\psi_{2/1}-\psi_{1/1})\Xi_{/1}
                 -(\psi_{2/11}-\psi_{1/11})\Xi$\cr\qquad\qquad
       $+(\psi_{1/2}-\psi_{2/2})\Xi_{/2}
       +2\psi_{1/1}\Xi_{/1}+2\psi_{1/2}\Xi_{/2}\}$\cr
       $\Cal{F}_{1,}{}^2\equiv\varepsilon\varrho\{
        -(\psi_{1/12}-\psi_{2/12})\Xi
      -2\psi_{1/2}\Xi_{/1}+2\psi_{2/1}\Xi_{/2}\}$
\cr}}}

\medbreak We now prove assertion (1). Let $P\in\text{int}(M)$. Let
$\varepsilon\psi_1(P)=\varepsilon\psi_2(P)=0$. We study monomials $\Xi_{/111}$ and 
  $\psi_{2/111}\Xi$
appearing in $\birdy{a_4^M(\cdot)}$. Let $\Cal{R}=E$ or let $\Cal{R}=R_{ijji}$. 
  We integrate by parts
to define
$\Cal{A}[\Cal{R}]$ by the identity:
$$\eqalign{
  &-\textstyle\frac1{12}\textstyle\int_Mg^{-1}\partial_1(gf\Xi)\Cal{R}d\nu_M
   =\frac1{360}\textstyle\int_M f\Cal{A}[\Cal{R}] d\nu_M;\text{ then}\cr
  &\textstyle-\frac12a_2^M(g^{-1}\partial_1(gf\Xi),\Delta)=
   (4\pi)^{-1}\frac1{360}\textstyle\int_M f \Cal{A}[6E+R_{ijji}]d\nu_M.}$$
We have $R_{ijji}\equiv-2\varepsilon\psi_{2/11}+...$. We compute:
\medbreak\centerline{\vbox{\offinterlineskip\halign{
\phantom{$\frac1{6_{\vrule height 6pt}}$}
 #\ \hfill\ &\hfill#\quad&\hfill#\quad&\hfill#
\cr
$\birdy{60E_{;ii}}$&$\equiv$&$-30\Xi_{/111}$&$-30\varepsilon\psi_{2/111}\Xi+...$\cr
$\birdy{60\Cal{F}_{1,i;i}}$&$\equiv$&$60\Xi_{/111}$&$-60\varepsilon\psi_{2/111}\Xi+...$\cr
$\birdy{c_9\Cal{G}_{1,ii;jj}}$&$\equiv$&$2c_9\Xi_{/111}$&$+2c_9\varepsilon\psi_{2/111}\Xi+...$\cr 
$\birdy{c_{10}\Cal{G}_{1,ij;ij}}$&$\equiv$&$2c_{10}\Xi_{/111}$&$+0c_{10}\varepsilon\psi_{2/111}\Xi+...$\cr
$\Cal{A}[6E]$&$\equiv$&$0\Xi_{/111}$&$+0\varepsilon\psi_{2/111}\Xi+...$\cr
$\Cal{A}[R_{ijji}]$&$\equiv$&$0\Xi_{/111}$&$-60\varepsilon\psi_{2/111}\Xi+...$\cr
}}}
\medbreak\noindent We use Lemma \erefa\ to relate the coefficients of $f\Xi_{/111}$ 
   and $f\psi_{2/111}\Xi$ and
establish the following relationships from which assertion (1) follows:
$$-30+60+2c_9+2c_{10}=0\text{ and }-30-60+2c_9=-60.$$

We now study the boundary terms. We pullback the Robin boundary operator
$$\Phi_\varrho^*(e^{-\varepsilon\psi_2}\partial_2+S)\equiv 
    e^{-\varepsilon\psi_{2/1}t\varrho\Xi}\{
     \Cal{B}-e^{-\varepsilon\psi_2}t\varrho\Xi_{/2}\partial_1
   +t\varrho\Xi( S\varepsilon\psi_{2/1}+ S_{/1})\}$$
to determine the tensors
$$T^1\equiv-e^{-\varepsilon\psi_2}\varrho\Xi_{/2}\text{ and }
      S_1\equiv\varrho\Xi(\varepsilon\psi_{2/1}S+S_{/1}).$$
We have $L_{11}\equiv-\varepsilon\psi_{1/2}$. 
We study the terms comprising $\birdy{a_4^{\partial M}(f,\Cal{D}_\varrho,
\Cal{B}_\varrho)}$.
At the point of the boundary in question, we suppose 
   $\varepsilon\psi_1(P)=\varepsilon\psi_2(P)=0$.
\BORK $\birdy{(-120^-,240^+) fE_{;m}}$\bork$\equiv(60^-,-120^+) f\{\Xi_{/12}
+(\varepsilon\psi_{1/12}+\varepsilon\psi_{2/12})\Xi+(\varepsilon\psi_{1/1}
+\varepsilon\psi_{2/1})\Xi_{/2}\}$,
\BORK $\birdy{120 fEL_{aa}}\equiv60 
          \varepsilon f \psi_{1/2}\Xi_{/1}$,
\BORK
$\birdy{720fSE}\equiv-360fS\{\Xi_1+\varepsilon(\psi_{1/1}
+\psi_{2/1})\Xi\}$,
\BORK $\birdy{e_3^\pm f\Cal{G}_{1,aa}L_{bb}}
            \equiv e_3^\pm f(2\Xi_{/1})(-\varepsilon\psi_{1/2})$,
\BORK $\birdy{e_4^\pm f\Cal{G}_{1,mm}L_{bb}}\equiv0$,
\BORK $\birdy{e_5^\pm f\Cal{G}_{1,ab}L_{ab}}\equiv
                               e_5^\pm f(2\Xi_{/1})(-\varepsilon\psi_{1/2})$,
\BORK $\birdy{e_6^\pm f\Cal{G}_{1,mm;m}}\equiv
   e_6^\pm f(2\varepsilon\psi_{2/12}\Xi+4\varepsilon\psi_{2/1}\Xi_{/2})$,
\BORK $\birdy{e_7^\pm f\Cal{G}_{1,aa;m}}
   \equiv e_7^\pm f\{2\Xi_{/12}+2\varepsilon\psi_{1/12}\Xi
    +2\varepsilon\psi_{1/1}\Xi_{/2}-2\varepsilon\psi_{2/1}\Xi_{/2}\}$,
\BORK $\birdy{e_8^\pm f\Cal{G}_{1,am;a}}
    \equiv e_8^\pm f\{-\varepsilon\psi_{2/1}\Xi_{/2}+\Xi_{/12}
     +\varepsilon\psi_{1/1}\Xi_{/2}-2\varepsilon\psi_{1/2}\Xi_{/1}\}$,
\BORK $\birdy{e_9^\pm f\Cal{F}_{1,m}}\equiv e_9^\pm f\{
       -(\varepsilon\psi_{1/12}-\varepsilon\psi_{2/12})\Xi
      -2\varepsilon\psi_{1/2}\Xi_{/1}+2\varepsilon\psi_{2/1}\Xi_{/2}\}$,
\BORK $\birdy{e_{12}^+ fS\Cal{G}_{1,aa}}\equiv 
      e_{12}^+ f\{2\Xi_{/1}S+2\varepsilon\psi_{1/1}\Xi S\}$,
\BORK $\birdy{e_{13}^+ fS\Cal{G}_{1,mm}}\equiv
      e_{13}^+ f\{2\varepsilon\psi_{2/1}\Xi S\}$,
\BORK $\birdy{e_{14}^+ fS_1}\equiv
      e_{14}^+ f\Xi\{\varepsilon\psi_{2/1}S+S_{/1}\}$,
\BORK $\birdy{e_{15}^+ fT_{a:a}}
    \equiv e_{15}^+ f(\varepsilon\psi_{2/1}\Xi_{/2}-\Xi_{/12}-
    \varepsilon\psi_{1/1}\Xi_{/2})$,
\BORK $\birdy{(\pm180)
f_{;m}E}\equiv\mp90
f_{;m}\{\Xi_{/1}+(\varepsilon\psi_{1/1}+
   \varepsilon\psi_{2/1})\Xi\}$,
\BORK $\birdy{e_{10}^\pm f_{;m}\Cal{G}_{1,aa}}
     \equiv e_{10}^\pm f_{;m}(
    2\Xi_{/1}+2\varepsilon\psi_{1/1}\Xi)$,
\BORK $\birdy{e_{11}^\pm f_{;m}\Cal{G}_{1,mm}}\equiv
     e_{11}^\pm f_{;m}2\varepsilon\psi_{2/1}\Xi$.
\medbreak\noindent We must also study the boundary terms comprising $-\frac12a_2^{\partial
M}(\cdot)$. As when studying $a_2^M$, we integrate by parts to define $\Cal{A}$ and compute:
\medbreak\BORK
$\Cal{A}[2fL_{aa}]
 \equiv-60\varepsilon f\psi_{1/12}\Xi$,
\BORK
$\Cal{A}[12fS]
 \equiv-360\{\Xi\varepsilon fS\psi_{2/1}
      -f\Xi S_{/1}\}$,
\BORK
$\Cal{A}[\pm3f_{;m}]
   \equiv\mp90\{(\varepsilon\psi_{1/12}+\varepsilon\psi_{2/12})
   f\Xi+2\varepsilon\psi_{2/1}(f_{;m}\Xi+f\Xi_{/2})\}$.
\medbreak\noindent
We established the following relations in Lemmas \crefb\ and \drefb:
$$e_3^\pm=30, e_4^\pm+e_5^\pm=-30,\ e_{14}^+-2e_9^+=60\text{ and }e_9^-=-30.$$
 We use Lemma \erefa\
to derive the following 
equations and complete the proof:\medbreak\centerline{\vrule\vbox{\offinterlineskip\hrule\halign{
\phantom{$\frac1{6_{\vrule height 4pt}}$} #\ \hfil&\ \hfill#\ 
\cr 
$(60^-,-120^+)+4e_6^\pm-2e_7^\pm-e_8^\pm+2e_9^\pm+e_{15}^\pm =\mp180$&
\quad[$f\varepsilon\psi_{2/1}\Xi_{/2}$]\cr\noalign{\hrule}
$(60^-,-120^+) +2e_6^\pm+e_9^\pm=\mp90$&[$f\varepsilon\psi_{2/12}\Xi$]\cr\noalign{\hrule}
$(60^-,-120^+) +2e_7^\pm-e_9^\pm=-60\mp90$&[$f\varepsilon\psi_{1/12}\Xi$]\cr\noalign{\hrule}
$(60^-,-120^+) +2e_7^\pm+e_8^\pm-e_{15}^\pm=0$&[$f\Xi_{/12}$]\cr
}}\vrule}
\centerline{\vrule\vbox{\offinterlineskip\hrule\halign{
\phantom{$\frac1{6_{\vrule height 6pt}}$} #\ \hfil&\ \hfill#&#\ \hfil&\ \hfill#
\cr 
$-2e_5^\pm-2e_8^\pm-2e_9^\pm =0$&[$f\varepsilon\psi_{1/2}\Xi_{/1}$]
   &\ \vrule \ $e_{14}^+ =360 $&$[fS_{/1}\Xi$]\cr\noalign{\hrule}
$-360+2e_{13}^++e_{14}^+ =-360 $&[$f\varepsilon\psi_{2/1}\Xi S$]
   & \ \vrule \ $-360+2e_{12}^+ =0$&$[f\Xi_{/1}S$]\cr\noalign{\hrule}
$\mp90+2e_{11}^\pm =\mp180 $&[$f_{;m}\varepsilon\psi_{2/1}\Xi$]
   &\ \vrule \ $\mp90 +2e_{10}^\pm =0$&[$f_{;m}\Xi_{/1}$]\cr
}\hrule}\vrule}\enddemo

\sethead\qctSF\head\S\qctSF\ Commuting operators\endhead
We conclude this paper by deriving a final functorial property.
The equations which can be derived using this property are compatible with the values for
the constants $c_i$ and $e_i^\pm$ previously computed; they are omitted in the interests of
brevity.

\setref\frefa
\proclaim{\frefa\ Lemma} Let $D$ be a self-adjoint static operator of Laplace type and let
$\Cal{B}$ be a static boundary condition. Let $Q$ be an auxiliary self-adjoint static
partial differential operator of order at most $2$ which commutes with $D$ and with
$\Cal{B}$. Then:
$$\birdy{a_n(f,D+2t\varrho Q,\Cal{B})}=
  \birdy{a_{n-2}(f,D+\varrho Q,\Cal{B})}.$$
\endproclaim
\subhead Remark\endsubhead If we take $D=Q$, then $D(\varrho)=(1+2t\varrho)D$. By Lemma \crefa,
$$\birdy{a_4(f,(1+2t\varrho)D,\Cal{B})}
   =\frac{2-m}2a_2(f,D,\Cal{B}).$$
On the other hand, clearly
   $a_n(f,(1+\varrho)D,\Cal{B})
    =(1+\varrho)^{(n-m)/2}a_n(f,D,\Cal{B})$.
Thus we may show that Lemma \frefa\ is compatible with Lemma \crefa\ in this special case
by computing:
$$\birdy{a_2(f,(1+\varrho)D,\Cal{B})
}=\frac{2-m}2a_2(f,D,\Cal{B})=\birdy{a_4(f,D+2t\varrho D,\Cal{B})}.$$

\demo{Proof} Let
$\Cal{K}_1(t)
:=(1-t^2\varrho Q)e^{-tD_\Cal{B}}$. Then
$\Cal{K}_1(0)$ is the identity operator and:
$$\eqalign{
   &(\partial_t+D+2t\varrho Q)(1-t^2\varrho Q)e^{-tD_\Cal{B}}\cr
  =&
   \{-2t\varrho Q-(1-t^2\varrho Q)D+D(1-t^2\varrho Q)+2t\varrho Q(1-t^2\varrho Q)\}
    e^{-tD_\Cal{B}}\cr
   =&-2t^3\varrho^2Q^2e^{-tD_\Cal{B}}.}$$
There exists a constant $C$ and an integer $\mu$ so that we have the estimate in a
suitable operator norm:
$$|-2t^3\varrho^2Q^2e^{-tD_{\Cal{B}}}|\le Ct^{-\mu}\varrho^2.$$
Thus since we are interested in the linear terms in $\varrho$, we may replace the
fundamental solution of the heat equation $\Cal{K}(t)$ for $D+2t\varrho Q$ by the
approximation
$(1-\varrho t^2Q)e^{-tD_{\Cal{B}}}$.
There is an asymptotic expansion of the form \cite{\refGilkey}:
$$\trace_{L^2}(fQe^{-tD_\Cal{B}})\sim\textstyle\sum_{n\ge0}
   t^{(n-m-2)/2}a_n(f,Q,D,\Cal{B}).$$
We equate coefficients of $t^{(n-m)/2}$ in the asymptotic expansions to see
$$\birdy{a_n(f,D+2t\varrho Q,\Cal{B})}=-a_{n-2}(f,Q,D,\Cal{B}).$$
Since $Q$ and $D$ commute and since $Q$ and $\Cal{B}$ commute, we complete the proof by
computing:
$$\eqalign{
&\textstyle\sum_{n\ge0}\birdy{a_n(f,D+\varrho Q,\Cal{B})}
t^{(n-m)/2}\sim
\birdy{\trace_{L^2}(fe^{-t((D+\varrho Q)_{\Cal{B}})})}\cr
&\qquad=\textstyle\trace_{L^2}(-tf Qe^{-tD_\Cal{B}})
\sim-\sum_{n\ge0}a_n(f,Q,D,\Cal{B})t^{(n-m)/2}\text{ so}\cr
&\birdy{a_n(f,D+\varrho Q,\Cal{B})}=-a_n(f,Q,D,\Cal{B}).\ \qed}$$\enddemo

\bigbreak\noindent{\bf Acknowledgements}: It is a pleasant task to thank G.
Steigelman and D. Vassilevich for helpful conversations about various
matters. The first and second author are
grateful for the support provided by MPI (Leipzig) during their visit there.
PG has been supported by a grant from the NSF (USA),
KK has been supported by the EPSRC under the grant number GR/M45726, and
JHP has been supported by a grant from the
Korea Research Foundation made in the program year of 1998.

\enddocument